\documentclass[slac_one]{revtex4}
\usepackage{graphicx}
\usepackage{fancyhdr}
\newcommand\met{\mathrm{E}_\mathrm{T}^\mathrm{miss}}
\newcommand\inb{\mathrm{nb}^{-1}}
\newcommand\pt{p_\mathrm{T}}
\newcommand\mot{m_\mathrm{T}}
\newcommand\meff{M_\mathrm{eff}}

\begin{document}

\title{Distributions for one-lepton SUSY Searches with the ATLAS Detector} 

%

\author{M-H. GENEST on behalf of the ATLAS Collaboration}
\affiliation{Ludwig-Maximilians-Universit\"at M\"unchen, Am Coulombwall 1, 85748 Garching, GERMANY}

\begin{abstract}
Using ATLAS data corresponding to $70 \pm 8\,\inb$ of integrated luminosity from the 7 TeV proton-proton collisions at the LHC, distributions of relevant supersymmetry-sensitive variables are shown for the final state containing jets, missing transverse momentum and one isolated electron or muon. With increased integrated luminosities, selections based on these distributions will be used in the search for supersymmetric particles: it is thus important to show that the Standard Model backgrounds to these searches are under good control.
\end{abstract}

\maketitle

\thispagestyle{fancy}


\section{INTRODUCTION}

If supersymmetry exists at the TeV scale and R-parity is conserved, the SUSY particles should be produced in pairs and decay to the lightest SUSY particle which would escape detection, thus leading to signatures containing jets, large missing transverse momentum and potentially one or more leptons. We present here a first comparison of the ATLAS data corresponding to $70 \pm 8\,\inb$ of integrated luminosity from the $\sqrt{s}=7$ TeV proton-proton collisions to Monte Carlo simulations for some of the most important kinematical variables that are expected to be sensitive for SUSY searches. A more detailed description of these results can be found in \cite{confnote}.

\section{THE ONE-LEPTON SUSY ANALYSIS}\label{sec:analysis}

In the one-lepton analysis, after a leptonic trigger requirement and a set of cleaning cuts to reject events containing jets which are consistent with calorimeter noise, cosmic rays or out-of-time energy deposits, the events are preselected by asking for at least two jets with transverse momentum $\pt>30$ GeV and one isolated lepton (electron or muon) with $\pt>20$ GeV. The signal region is then defined by applying two further cuts: $\met>30$ GeV and $\mot>100$ GeV, where $\met$ is the missing transverse momentum, calculated as the opposite of the vector sum of the transverse energies of all three-dimensional topological clusters in the calorimeter plus the transverse momenta of the selected well-isolated muons in the analysis and $\mot$ is the transverse mass of the lepton and $\met$ defined as $m_\mathrm{T}^2 \equiv 2|{\bf p}_\mathrm{T}^{\ell}||\met| - 2{\bf p}_\mathrm{T}^{\ell} \cdot \vec \met$.

The data is compared to the full-detector GEANT4 simulation which is reconstructed using the same algorithms as for the data. The Standard Model background processes considered in this analysis are QCD (PYTHIA), W/Z+jets (ALPGEN + HERWIG + Jimmy) and $t\bar{t}$ (MC@NLO + HERWIG + Jimmy). The PYTHIA QCD predictions were compared to a set of ALPGEN QCD samples; the differences were found to be well within the experimental uncertainties for the kinematic region explored. The QCD and W+jets backgrounds are normalized to the data in control regions defined as $\met<40$ GeV and $\mot<40$ GeV for the QCD background and $30<\met<50$ and $40<\mot<80$ GeV for the W+jets background. As an example, the SU4 supersymmetric point (ISAJET + HERWIG + PROSPINO) is also shown in the plots with its cross section multiplied by 10; SU4 is a low-mass benchmark point close to the Tevatron limits and is defined as $m_0=200$ GeV, $m_{1/2}=160$ GeV, $A_0=-400$ GeV, $tan\beta=10$ and $\mu>0$.

The most important sources of systematic uncertainties are considered: the uncertainty on the jet energy scale (which varies from 7-10$\%$ as a function of the jet $\pt$ and $\eta$), the uncertainty on the W+jets and QCD normalizations (50$\%$), the uncertainty on the Z+jets normalization (60$\%$) and the uncertainty on the luminosity (11$\%$).

\begin{figure}[htb]
\centering
\includegraphics[height=2.5in]{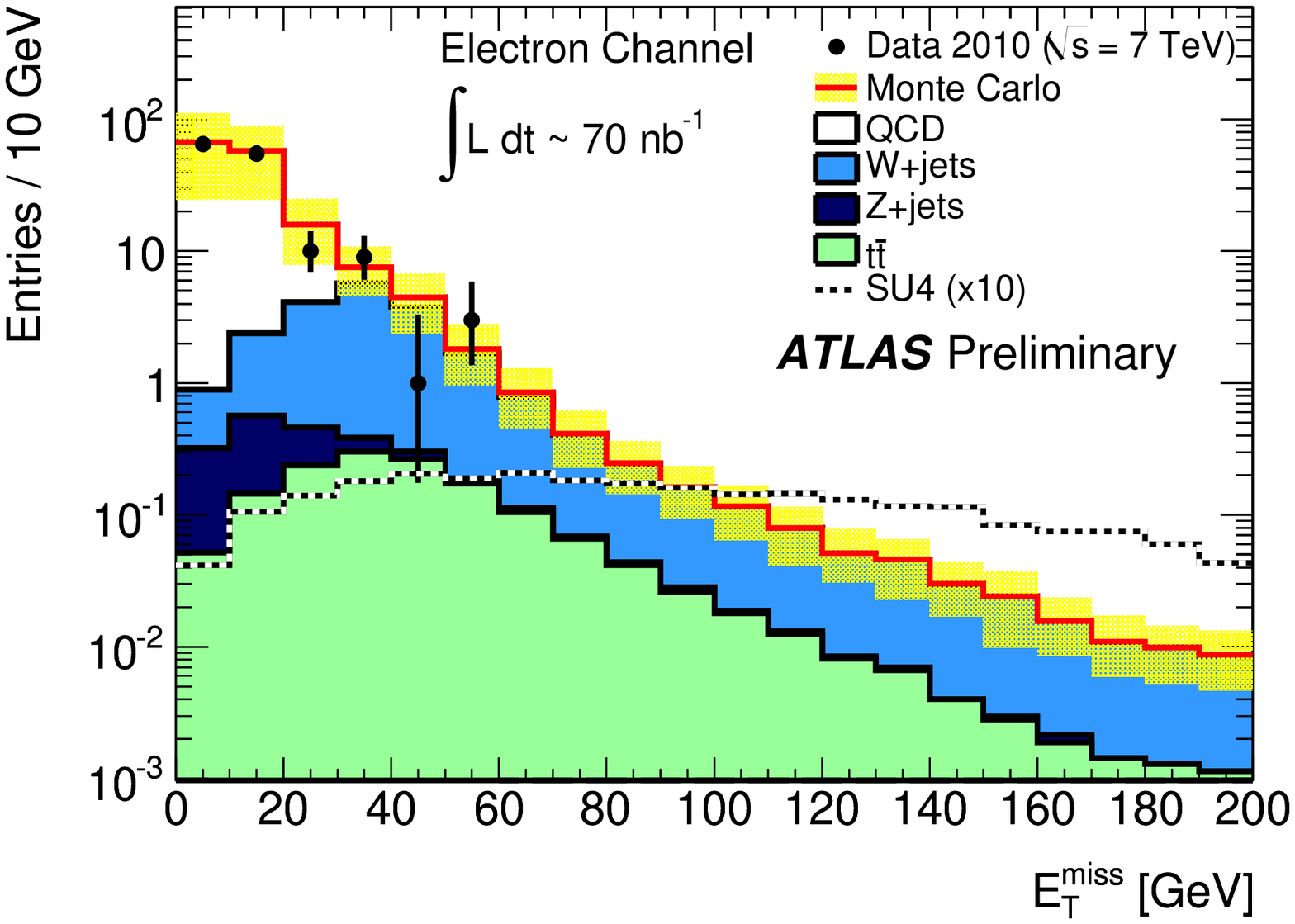}
\includegraphics[height=2.5in]{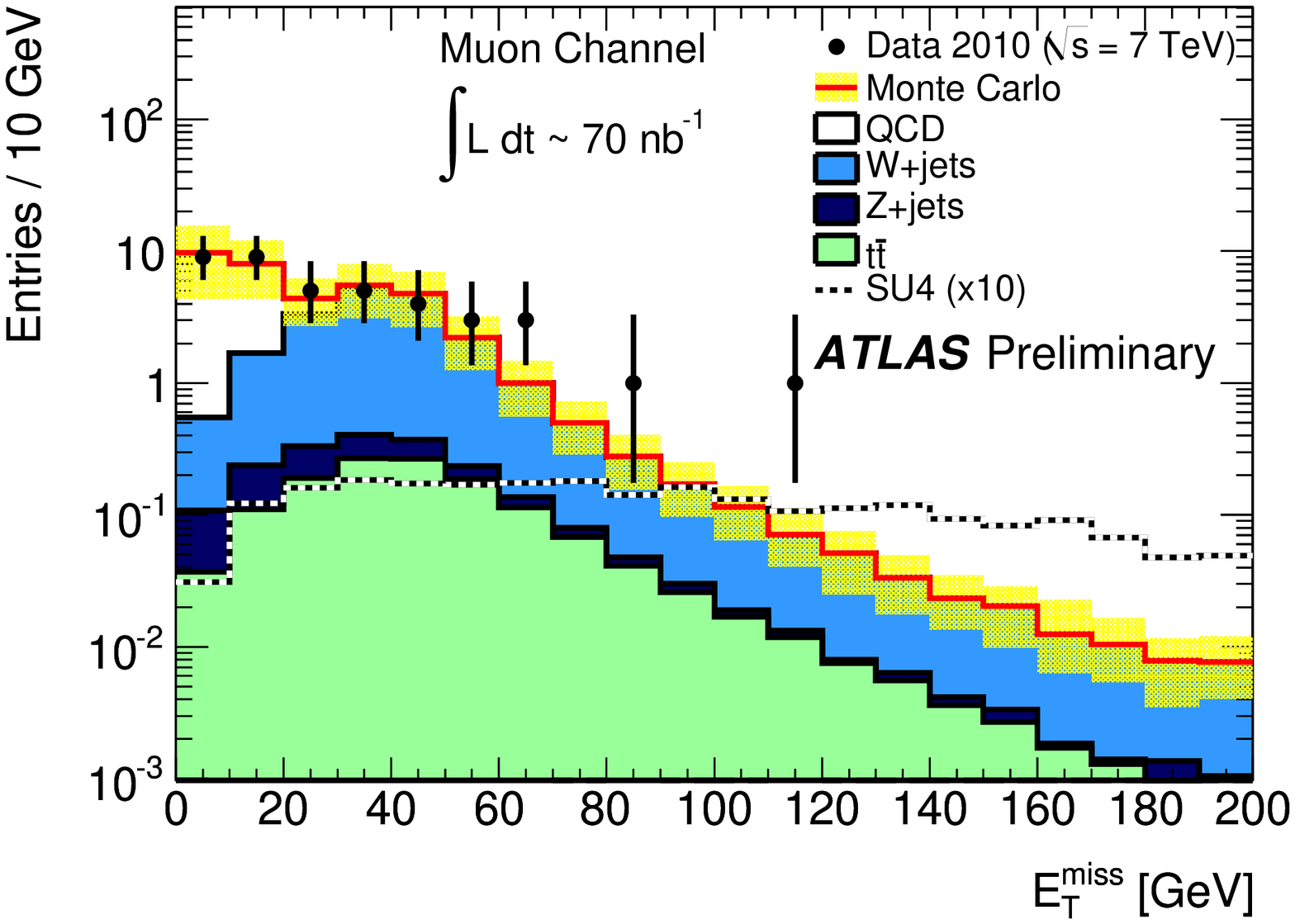}
\caption{$\met$ distribution after the preselection for the electron (left) and muon (right) channels. The statistical and systematic uncertainties on the Monte Carlo prediction, added in quadrature, are shown as a yellow band on the plots.
}
\label{fig:met}
\end{figure}

\begin{figure}[htb]
\centering
\includegraphics[height=2.5in]{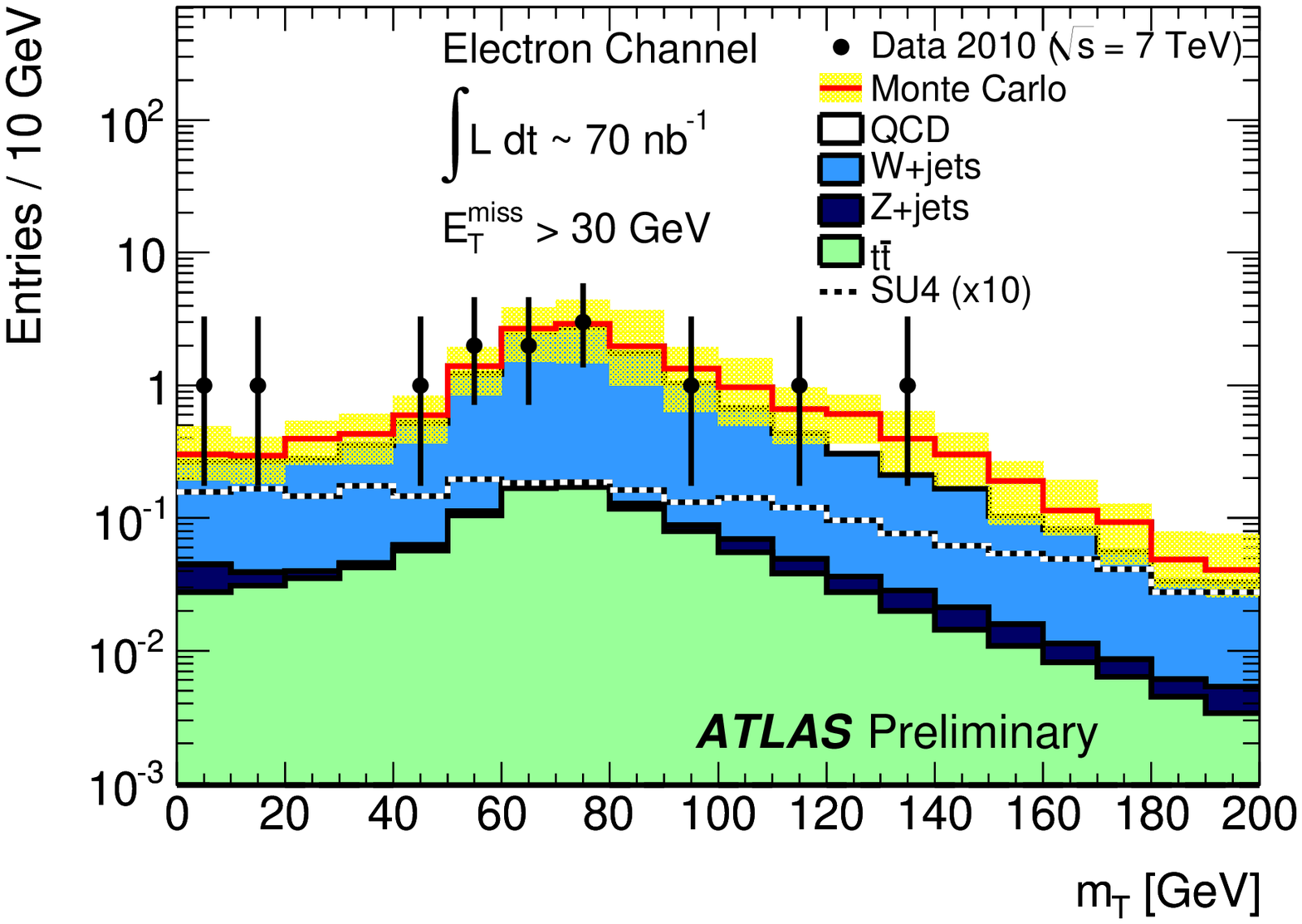}
\includegraphics[height=2.5in]{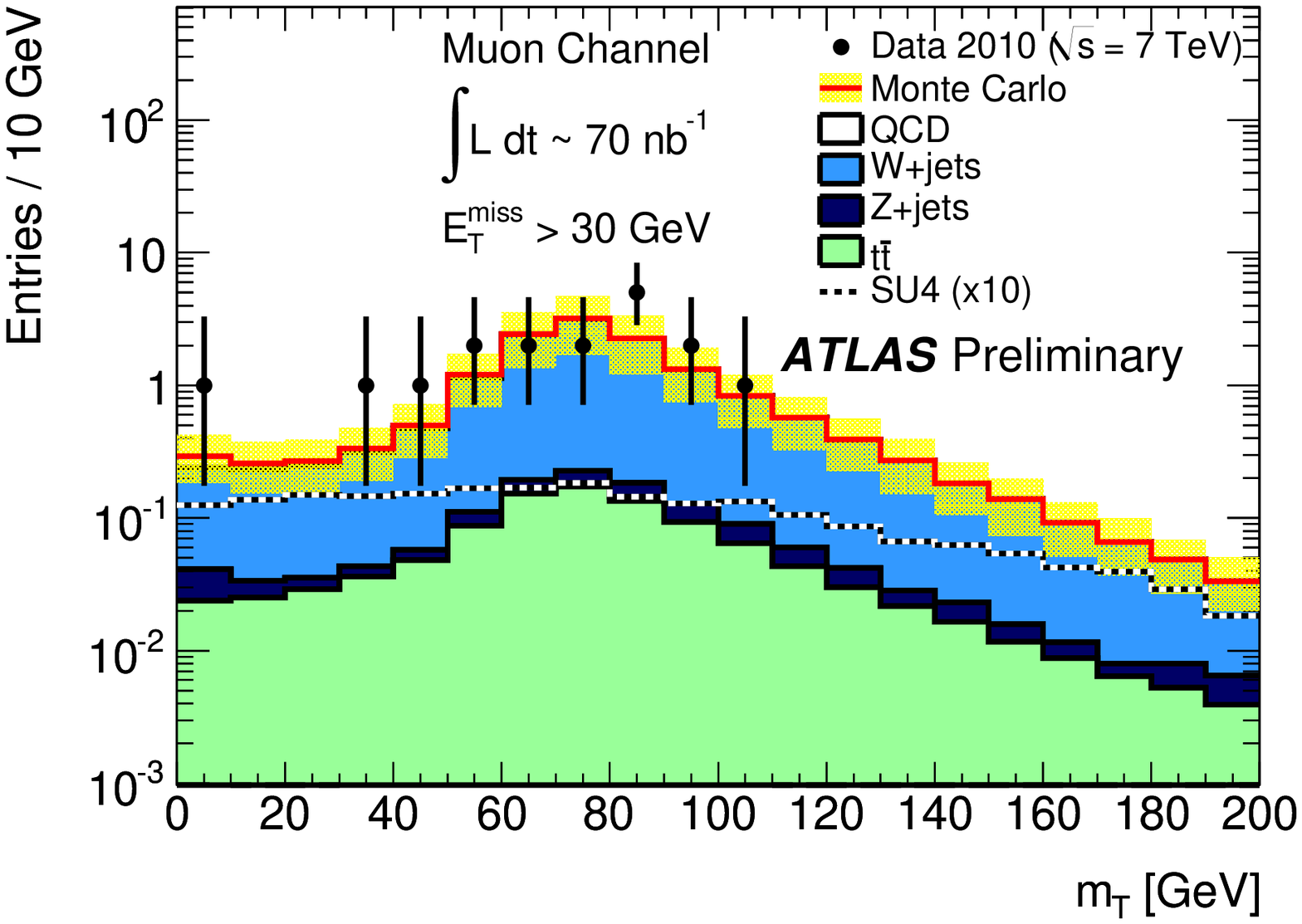}
\caption{$\mot$ distribution after the cut on $\met$ for the electron (left) and muon (right) channels.}
\label{fig:mot}
\end{figure}

\begin{figure}[htb]
\centering
\includegraphics[height=2.5in]{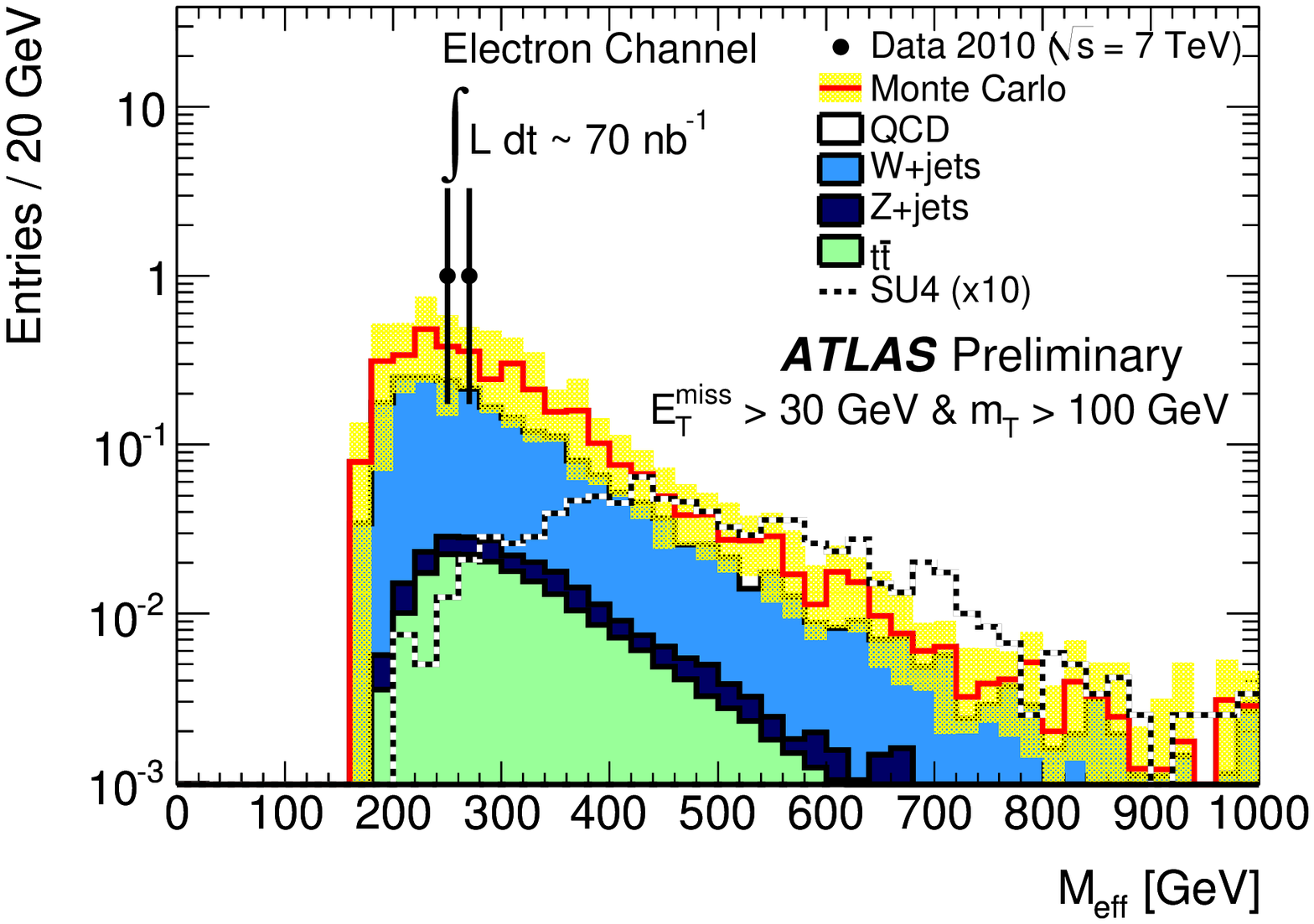}
\includegraphics[height=2.5in]{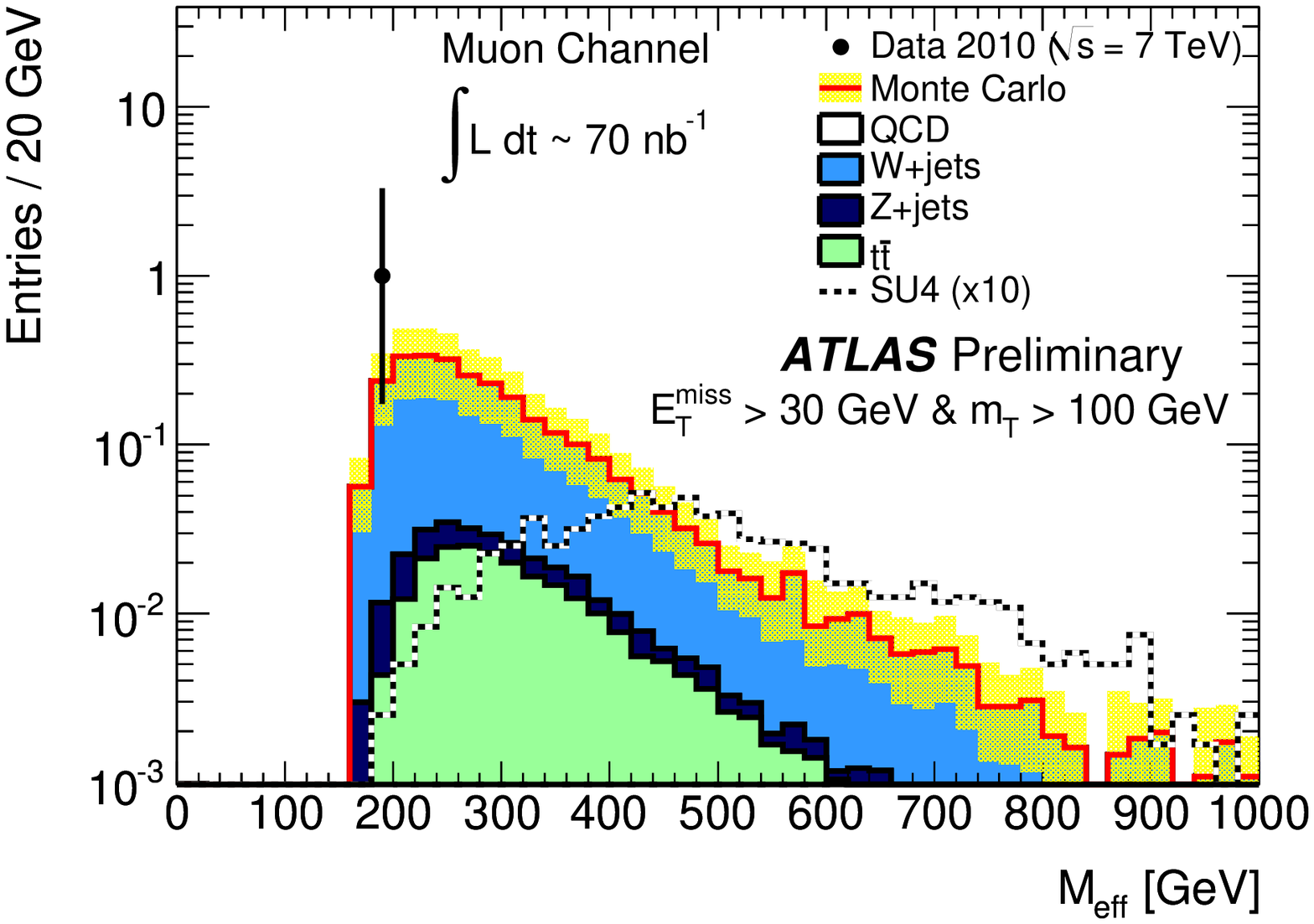}
\caption{$\meff$ distribution after the cuts on $\met$ and $\mot$ for the electron (left) and muon (right) channels.}
\label{fig:meff}
\end{figure}

The results are shown in Figures \ref{fig:met}-\ref{fig:meff}. Figure \ref{fig:met} shows $\met$ after the preselection for the electron and muon channels: there is reasonable agreement between the data and the Monte Carlo. While the low $\met$ region is dominated by the QCD background, the W+jets background dominates at higher values; the supersymmetry model would yield even higher $\met$ values. After applying the $\met>30$ GeV cut, the $\mot$ distributions for both channels are shown in Fig. \ref{fig:mot} and exhibit good agreement between the data and Monte Carlo. Finally, Fig. \ref{fig:meff} shows the effective mass distribution ($\meff$, defined as  $\meff \equiv \sum_{i=1}^{2} p_\mathrm{T}^{{jet},i} + p_\mathrm{T}^{{lep}} + \met)$ for the signal region, i.e. after a further cut on $\mot>100$ GeV. The number of events found in the signal region is consistent with the expectations. The expected number of events is compared with the data at different stages of the analysis in Table \ref{tab:results}.

\begin{table}\renewcommand\arraystretch{1.2}
        \begin{center}
                        \begin{tabular}{|l|rr|rr|}
                                \hline
                                & \multicolumn{2}{c|}{Electron channel} & \multicolumn{2}{c|}{Muon channel} \\
                                Selection   & Data & Monte Carlo & Data & Monte Carlo \\\hline\hline
                                \begin{minipage}{5cm}
			         \begin{flushleft}
                                        $p_\mathrm{T}(\ell)>20\,$GeV$ \ \cap$ \\
                                        $\geq2$ jets with $p_\mathrm{T}>30\,$GeV
				 \end{flushleft}
                               \end{minipage}
                                &   143      &  157 $\pm$ 85 &  40  & 37 $\pm$ 14 \\ \hline
                                $\cap~ \met > 30\,$GeV & 13 & 16 $\pm$ 7 & 17 & 15 $\pm$ 7 \\ \hline
                                $\cap~ \mot > 100\,$GeV & 2 & 3.6 $\pm$ 1.6 & 1 & 2.8 $\pm$ 1.2 \\\hline
                        \end{tabular}
                \caption{\label{tab:results} Number of events observed and predicted at several stages of the single lepton selection.  As described in Section~\ref{sec:analysis}, the Monte Carlo predictions have been normalised to the data in control regions which overlap all but the final selection.  }
        \end{center}
\end{table}
\section{CONCLUSION}
The first $70 \pm 8\,\inb$ of integrated luminosity collected with the ATLAS detector are analysed in an early search for new physics in the channel containing jets, missing transverse momentum and one lepton (electron or muon). The measurements are compared to simulations of the expected Standard Model background and generally show agreement with these expectations.

\begin{acknowledgments}
The author would like to acknowledge support by the DFG cluster of excellence "Origin and Structure of the Universe" (www.universe-cluster.de).
\end{acknowledgments}

\end{document}